\documentclass[11pt,superscriptaddress,prd,nofootinbib,showkeys]{revtex4}

\usepackage{amssymb,amsmath}
\usepackage{epsfig}

\newcommand{\beq}{\begin{equation}}
\newcommand{\eeq}{\end{equation}}
\newcommand{\baq}{\begin{eqnarray}}
\newcommand{\eaq}{\end{eqnarray}}
\newcommand{\mc}[1]{\mathcal{#1}}

\def\Tdot#1{{{#1}^{\hbox{.}}}}

\def\cov{\nabla} 
\def\lie{{\cal L}} 
\def\cv{e} 
\def\lv{{\tilde e}}

\begin{document}

\title{Covariant generalization of cosmological perturbation theory}

\author{Kari Enqvist}
\email{kari.enqvist@helsinki.fi}
\affiliation{Helsinki Institute of Physics, P.O. Box 64, FIN-00014 University
  of Helsinki, Finland}
\affiliation{Department of Physical Sciences, P.O. Box 64, FIN-00014
University of Helsinki, Finland}
\author{Janne H\"{o}gdahl}
\email{janne.hogdahl@helsinki.fi}
\affiliation{Helsinki Institute of Physics, P.O. Box 64, FIN-00014 University of Helsinki, Finland}
\author{Sami Nurmi}
\email{sami.nurmi@helsinki.fi}
\affiliation{Department of Physical Sciences, P.O. Box 64, FIN-00014
University of Helsinki, Finland}
\author{Filippo Vernizzi}
\email{vernizzi@ictp.it}
\affiliation{Helsinki Institute of Physics, P.O. Box 64, FIN-00014 University of Helsinki, Finland}
\affiliation{Abdus Salam ICTP, Strada Costiera 11, 34100 Trieste, Italy}

\begin{abstract}
We present an approach to cosmological perturbations based
on a covariant perturbative expansion between two worldlines in the
real inhomogeneous universe. As an application, at an arbitrary order we
define an exact scalar quantity which
describes the inhomogeneities in the number of e-folds on uniform
density hypersurfaces and which is conserved on all scales for a
barotropic ideal fluid. We derive a compact form for its
conservation equation at all orders and assign it a simple physical
interpretation. To make a comparison with the standard perturbation theory, we
develop a method to construct gauge-invariant quantities in a
coordinate system at arbitrary order, which we apply to derive the form of
the $n$-th order perturbation in the number of e-folds
on uniform density hypersurfaces and its exact evolution equation.
On large scales, this provides the gauge-invariant expression for
the curvature perturbation on uniform density hypersurfaces and its
evolution equation at any order.
\end{abstract}

\preprint{HIP-2006-48/TH}


\maketitle

\section{Introduction}

Relativistic perturbation theory is an extremely useful tool for
studying primordial inhomogeneities in cosmology and interpreting
cosmological data such as the Cosmic Microwave Background (CMB)
temperature anisotropies. The {\em linear} perturbation theory has
been developed to a very high degree of sophistication
\cite{Lifshitz:1963ps,Bardeen:1980kt,Kodama:1985bj,Mukhanov:1990me}
and due to the smallness of the CMB temperature fluctuations
usually provides an excellent approximation of the evolution of
cosmological inhomogeneities. However, in order to reliably
compare the theory with the high accuracy of the present and
future observations, it has become important to study
perturbations {\em beyond} the linear order, and recently there
has been a flourishing activity around this subject. The foundation of
second and higher order perturbation theory has
been given in \cite{Bruni,secondorder}, although already at second
order the coordinate approach becomes computationally challenging (see for
example \cite{Bartolo,Noh:2004bc,Nakamura}).

Instead of resorting to higher orders, one can alternatively use
non-perturbative methods to study the nonlinear evolution of
cosmological perturbations. Most of these methods
\cite{Salopek:1990jq,Comer:1994np,Sasaki:1998ug,Lyth:2003im,Rigopoulos:2003ak,proof}
are based on the so-called long wavelength approximation, which
restricts their validity to super-Hubble scales.

A non-perturbative and covariant way to deal with relativistic
inhomogeneities has been proposed by Ellis and Bruni
\cite{Ellis:1989jt} and thereafter employed mainly in its
linearized version to study cosmological perturbations as an
alternative to the coordinate approach \cite{Bruni:1992dg}. There
one defines perturbations of scalar quantities as covectors that
vanish in a spatially homogenous and isotropic
Friedmann-Lema\^{\i}tre-Robertson-Walker (FLRW) universe. The
advantage of the covariant formalism is that it directly deals
with geometrical quantities that are not hampered by the gauge
problems of the standard perturbation theory.

Recently, the covariant formalism was revived in
\cite{Langlois:2005ii,Langlois:2006iq,Langlois:2006vv}
as a convenient way to derive exact equations describing the
nonlinear evolution of perturbations, without resorting to any
approximations. In \cite{Langlois:2005ii} it was
shown that, for a barotropic ideal fluid, it is possible to define
a covector which is conserved {\em exactly} and on all scales in
the Lie derivation along the fluid flow.

The existence of a conserved quantity in cosmology is of paramount
importance as it allows one to establish a connection between
perturbations at different epochs without solving for the detailed
evolution of the universe. The result of
\cite{Langlois:2005ii} generalizes the approximate
large-scale conservation of the so-called curvature perturbation
on uniform density hypersurfaces in perturbation theory. This
quantity was introduced in linear theory in
\cite{Bardeen:1980kt,Bardeen}, rederived using the continuity
equation alone in \cite{Hwang:1991aj,Wands:2000dp} and extended to
second order in the perturbations in \cite{Malik:2003mv}.
Nonlinear generalizations for the curvature perturbation have also
been derived using the long wavelength approximation in
\cite{Lyth:2003im,Rigopoulos:2003ak,proof}.

In the present paper, we present a covariant approach for the
study of nonlinear perturbations and their conservation properties
based on a perturbative expansion similar to the standard
coordinate-based perturbation theory. However, in the coordinate
approach a perturbation is defined as the difference between the
values that a quantity (a scalar or a tensor) takes in the real
inhomogeneous universe and in a fictitious ideal background,
whereas in our approach a perturbation is defined geometrically in
the real inhomogeneous universe as the difference between the
values of a quantity measured by two observers on different
worldlines. Thus, the terms in our covariant expansion are tensors
in the real universe,
not coordinate-based perturbations around an ideal homogeneous universe.
Our definition
also implies that the perturbation of a tensor of a given type is
described by a tensor of the same type. Thus, in contrast to the
approach of \cite{Ellis:1989jt}, the perturbations of scalar
fields are described by scalars and not by covectors.

As in the standard coordinate approach, our perturbative expansion
allows us to write a hierarchy of equations which, order by order,
are coupled with those of lower order. However, being covariantly
defined in the  real universe, our quantities are automatically
gauge-independent, thus avoiding the complications of the gauge
issue of the standard perturbation theory.

To demonstrate the efficiency of our approach, we apply it to the
continuity equation of an ideal fluid. At each order in
covariant perturbations, we define a scalar quantity $\zeta_{(n)}$
which generalizes, on large scales, the curvature perturbation on
uniform density hypersurfaces at the same order. We derive an
exact evolution equation for $\zeta_{(n)}$, showing that at
arbitrary order it is conserved for a barotropic ideal fluid. This
equation manifestly mimics the perturbative large-scale evolution
equation of the curvature perturbation on uniform density
hypersurfaces.

This paper is organized as follows. In Sec.~\ref{sec:covariant} we
briefly review the covariant formalism for nonlinear cosmological
perturbations. In Sec.~\ref{sec:newcovariant} we define our
covariant perturbative expansion and  apply it to the continuity
equation. Section \ref{sec:coordinates} is devoted to the
comparison between our covariant perturbative expansion with the
standard coordinate-based approach. Finally, in
Sec.~\ref{sec:conclusion} we draw our conclusions.

\section{Covariant formalism}
\label{sec:covariant}

In this section we briefly review the basic ideas of the covariant
approach to perturbations proposed in \cite{Ellis:1989jt} and
recently used to study the conservation and evolution of nonlinear
relativistic perturbations for a perfect fluid
\cite{Langlois:2005ii}, for dissipative
interacting fluids \cite{Langlois:2006iq} and for scalar fields
\cite{Langlois:2006vv}.

We consider a universe dominated by an ideal fluid characterized
by a four-velocity $u^{a}= d x^{a} / d \tau$ ($u_{a} u^{a} =-1$),
where $\tau$ is the proper time of an observer comoving with the
fluid. The energy-momentum tensor of the fluid is
\beq T^{a}_{\ b}=\rho u^{a} u_{b}+ph^{a}_{\ b},
\label{emt}
\eeq
where $h^{a}_{\ b}$ is the spatial projection
tensor orthogonal to the four-velocity $u^{a}$,
\beq
h_{ab}\equiv g_{ab}+u_{a} u_{b}, \quad \quad (h^{a}_{\ c} h^{c}_{\
b}=h^{a}_{\ b}, \quad h^{a}_{\ b}u^{b}=0).
\eeq

In the covariant approach, the deviations from a spatially
homogeneous and isotropic FLRW universe are described by tensor
fields defined in the real inhomogeneous universe and vanishing in
an FLRW spacetime. The inhomogeneities of scalar quantities that
do not vanish in the background can be described in a simple way
by considering the {\em spatial projections} of their covariant
derivatives \cite{Ellis:1989jt}. For any scalar $f$ one can define
the projected gradient
\beq
D_{a} f \equiv h_{a}^{\ b}
\nabla_{b} f
\label{spatial_gradient},
\eeq
which vanishes in the FLRW background and is thus interpreted as a
perturbation. This definition is purely geometrical and depends
only on the four-velocity $u^a$. Since the covector $D_{a}f$ is
defined in the real inhomogeneous universe,
it can be
conveniently employed to describe nonlinear perturbations without
having to rely on 
a perturbative expansion around an ideal homogeneous background.

In \cite{Langlois:2005ii}, the covariant approach
was used to define conserved nonlinear perturbations
from the continuity equation, obtained by contracting the conservation
equation of the energy-momentum tensor,
\beq \nabla_{b}T^{b}_{\
a}=0,
\eeq
with the fluid four-velocity $u^a$. For an ideal fluid, the
continuity equation reads
\beq
\dot \rho + 3 \dot
\alpha (\rho +p)=0, \label{continuity}
\eeq
where $\alpha$ is the local number of e-folds defined as the
integral of the volume expansion along a worldline,
\beq
\label{alpha} \alpha\equiv
\frac{1}{3} \int \nabla_{a} u^{a} d\tau, \qquad (3 \dot \alpha =
\nabla_{a} u^{a}),
\eeq
and the dot denotes the covariant derivative projected along
$u^{a}$,
\beq
\dot{\alpha} \equiv \cov_u \alpha.
\eeq
Here, for convenience, we have adopted the short notation
\beq
\cov_v \equiv v^a \nabla_a \label{cov}
\eeq
for the covariant derivative along a generic vector $v^a$, that we
shall use hereafter. We remind here also the expression for the
Lie derivative along a vector $v^a$, which will be useful in the
following. For a scalar quantity $f$, this is equivalent to the
covariant derivative along $v^a$, i.e.,
\beq
\lie_v f = \cov_v f.
\eeq
For a vector $w^a$ it is equivalent to the commutator between
$v^a$ and $w^a$,
\beq
\lie_v w^a = [v,w]^a \equiv \cov_v w^a -\cov_w v^a,
\eeq
while for a covector $w_a$ it is given by
\beq
\lie_v w_a = \cov_v w_a + w_b \nabla_a v^b.
\eeq

In order to ``extract'' an evolution equation for covariant
perturbations from Eq.~(\ref{continuity}), one can simply take the
spatially projected gradient $D_{a}$ of this equation and invert
the time derivative and the spatial gradient. One
can define a covector
 \cite{Langlois:2005ii},
\beq \zeta_{a}
\equiv D_{a} \alpha -\frac{\dot \alpha}{\dot \rho} D_{a} \rho,
\label{zeta}
\eeq
that satisfies a remarkably simple conservation equation in the
Lie derivative along $u^a$,
\beq
 \lie_u \zeta_a = \frac{3 \dot \alpha^2}{\dot \rho}
\left( D_{a} p -\frac{\dot p}{\dot \rho} D_{a} \rho \right).
\label{zeta_a_conservation}
\eeq

Equation (\ref{zeta_a_conservation}) is fully nonlinear and exact
on all scales. For a barotropic fluid, i.e., when the pressure of
the fluid is a unique local function of the energy density,
$p=p(\rho)$, its right hand side vanishes and $\zeta_a$ is exactly
conserved  under Lie derivation along the worldlines of comoving
observers. When expanded to first order in the standard
perturbation theory, $\zeta_{a}$ reduces {\em on large scales} to
the spatial gradient of the first-order curvature perturbation on
uniform density hypersurfaces, usually denoted by $\zeta$. Furthermore, on
large scales Eq.~(\ref{zeta_a_conservation}) mimics the evolution equation
of $\zeta$.

\section{Covariant perturbation theory}
\label{sec:newcovariant}

\subsection{Covariant perturbations}
\label{sec:definition}

Here we define a covariant perturbative expansion in the real
inhomogeneous universe, which allows us to introduce the concept
of $n$-th order covariant perturbations. We first consider the
perturbations of a scalar quantity $f$. Later, we will generalize
our definitions to a tensor field.

In an inhomogeneous universe, two comoving observers living on
different worldlines and measuring the same proper time $\tau$
will in general observe different values of $f$ whereas in an FLRW
universe they would measure the same (homogeneous) value. Thus,
one can describe the inhomogeneities of $f$ by considering the
difference between the values of $f$ measured by two neighboring
comoving observers.

To characterize the separation between these observers, we introduce
a connecting vector $\cv^{a}\equiv dx^{a}/d\lambda$ that commutes
with the fluid four-velocity \cite{ellis},
\beq
[ u, \cv ]^{a} =0, \qquad (\Leftrightarrow  \quad \lie_u \cv^a =
0). \label{commutation_rel}
\eeq
As shown in Fig.~1, for each value of the proper time $\tau$, the
connecting vector $\cv^a$ defines an integral curve, parameterized
by $\lambda$, linking observers that measure the same proper time.
These observers define a continuous family of worldlines and
Eq.~(\ref{commutation_rel}) guarantees that the parameters
$\tau,\lambda$ can be used as (timelike and spacelike,
respectively) coordinates on this two-dimensional hypersurface.
\begin{figure}
\begin{center}
\includegraphics[height=20pc]{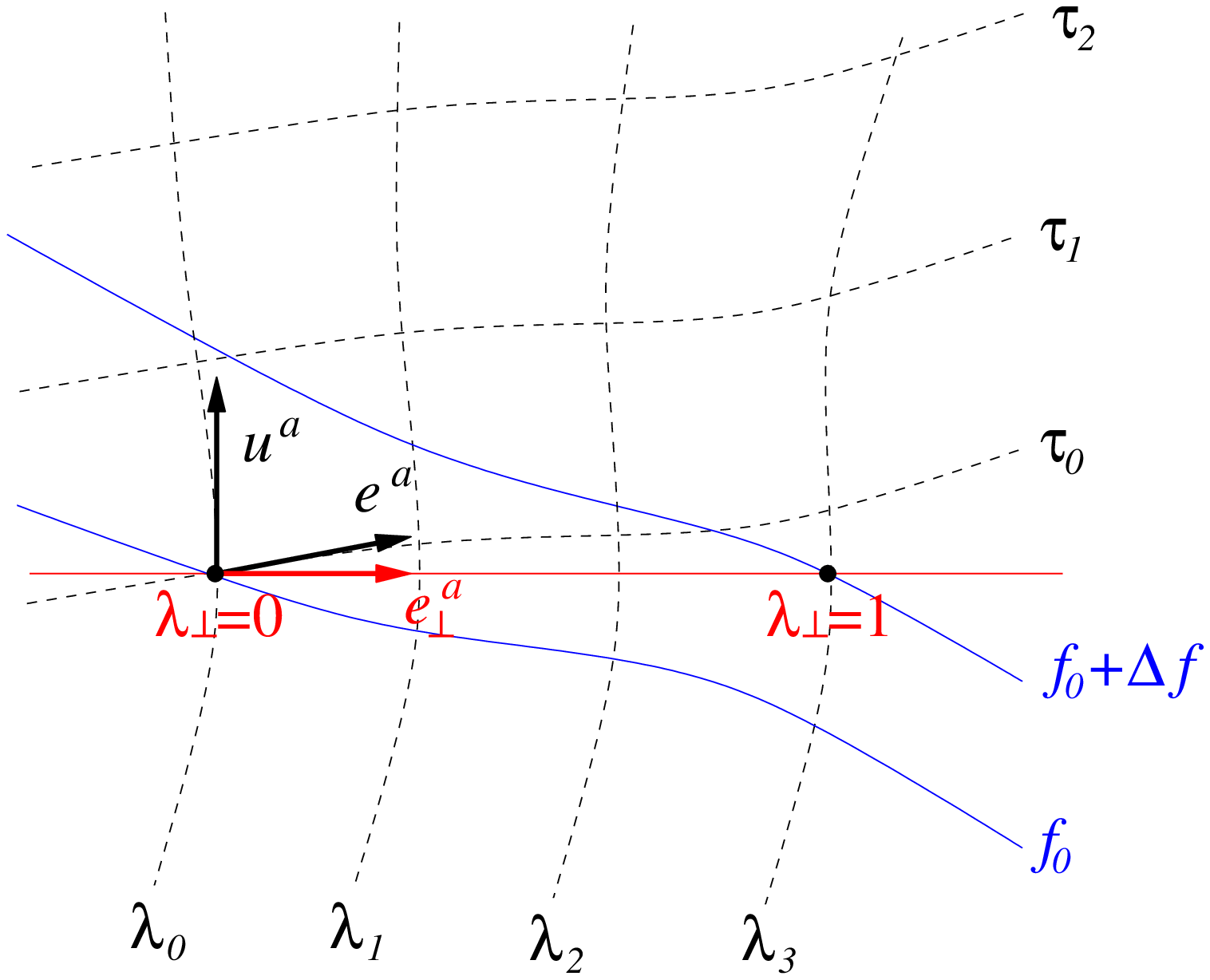}
\caption{Geometric definition of the covariant perturbation
$\Delta f_{\lambda_\perp}$} on the two-dimensional hypersurface
with the coordinates $\{\tau,\lambda\}$.\label{fig1}
\end{center}
\end{figure}

The connecting vector $\cv^a$ is not, in general, orthogonal to the
fluid four-velocity $u^{a}$, but can be decomposed into a
longitudinal part, parallel to $u^a$, and a relative position vector
$\cv_\perp^{a}$, orthogonal to $u^{a}$ \cite{ellis}, as
\beq
\cv^{a} \equiv \cv^{a}_\perp + \cv_\parallel
u^{a}\label{connecting}, \qquad (\cv_{\perp}^{a} = h^{a}_{\
b}\cv^{b},~\cv_{||} = -u^{a}\cv_{a}).
\eeq
The relative position vector $\cv_\perp^a$ defines another
integral curve, parameterized by $\lambda_\perp$,  which
characterizes the separation between neighboring observers as
measured in their rest frame.

We now define the perturbation of $f$ by comparing the values
measured by observers on the integral curve of $\cv_\perp^a$. For
convenience, we choose the parametrization of the curve such that
these observers are at $\lambda_\perp=0$ and $\lambda_\perp=1$,
respectively:\footnote{For brevity, we denote the point on a
manifold by the corresponding value of the curve parameter, i.e.,
we write $ f(\lambda_\perp)\equiv f(\sigma(\lambda_\perp))$ where
$\sigma$ is the integral curve of $\cv^{a}_\perp$.}
\beq
\Delta f\equiv f(1)-f(0).
\eeq
This  can be taylor-expanded along the integral curve of
$\cv_\perp^{a}$ as,
\beq
\Delta f=\sum_{n=1}^{\infty} \frac{1}{n!} \left.\cov_{\cv_\perp}^n
f\right|_{0}, \label{expansion_f}
\eeq
where $\left. \right|_{0}$ denotes the restriction to
$\lambda_\perp=0$. Each term in the expansion (\ref{expansion_f})
vanishes in the FLRW background and thus we define the $n$-th
order term in Eq.~(\ref{expansion_f}) as the $n$-th order {\em
covariant} perturbation of $f$.

The first-order term of the expansion (\ref{expansion_f}), i.e.,
the covariant first-order perturbation ${\cov_{\cv_\perp}} f$, has
actually been introduced already in the appendix of
\cite{Ellis:1989jt} as a scalar analogue of the spatially
projected gradient $D_a f$ employed in the standard covariant
approach. Indeed, it can be written as
\beq \cov_{\cv_\perp} f = \cv^a D_a f.
\label{rel_1}
\eeq
In contrast, the higher order covariant perturbations in (\ref{expansion_f})
have not been discussed previously in the literature.

The definition of the covariant $n$-th order perturbation of a
scalar can be easily generalized to an arbitrary tensor field $T$ living in
the inhomogeneous real spacetime, which we denote by ${\cal M}$,
by defining a diffeomorphism between $\lambda_\perp=0$ and
$\lambda_\perp=1$. This can be accomplished by using the flow
$\Phi:\mathbb{R} \times {\cal M} \to {\cal M}$, generated by the
vector field $\cv_\perp^{a}$. The diffeomorphism $\Phi$ defines a
pullback $\Phi^{*}_{\lambda_\perp}$ which maps a tensor $T$ at
$\lambda_\perp$ into a tensor $\Phi^{*}_{\lambda_\perp}T$ at
$\lambda_\perp=0$. The perturbation of the tensor field $T$ can
then be defined as
\beq \Delta
T \equiv \Phi^*_{1} \left.T \right|_{0} - \left.T\right|_{0}
\label{deltaf_def}.
\eeq
In an FLRW universe any tensor field $T$ is invariant under
$\Phi^{*}_{\lambda_\perp}$ because this corresponds to a
transformation of the coordinates on the homogeneous
spatial hypersurfaces. Therefore $\Delta T$ describes a perturbation around the
FLRW background and, as in the scalar case, it can be
taylor-expanded as
\beq
\Delta T = \sum_{n=1}^{\infty} \frac{1}{n!} \left.
\lie_{\cv_\perp}^n T \right|_{0}. \label{expansion_ff}
\eeq
The n-th order term $\lie_{\cv_\perp}^n \left.T \right|_0$ in this
expansion is the $n$-th order covariant perturbation of the tensor
field $T$. If $T$ is a scalar field, the Lie derivative is just
the directional derivative and we recover the expression
(\ref{expansion_f}).

The perturbation (\ref{deltaf_def}) and the perturbative expansion
(\ref{expansion_ff}) have been defined here in analogy to the
standard coordinate-based perturbation theory \cite{Bruni}.
However, instead of living in an ideal background spacetime,
$\Delta T$ and each term in its expansion are geometrically
defined quantities in the real inhomogeneous universe.
The $n$-th order covariant perturbations $\lie^n_{e_{\perp}}T$ measure
how the tensor field $T$ changes in the direction of
$e_{\perp}^{a}$. Therefore they depend on the choice of $\cv_{\perp}^{a}$ but
this is not a source of ambiquity since $\cv_{\perp}^{a}$ is a vector field
in the real universe and can be given a physical meaning.

\subsection{Evolution of perturbations}
\label{sec:evolution}

Now we study the evolution of perturbations employing our
covariant perturbative expansion for scalar quantities and
applying it to the continuity equation. Using this perturbative
expansion we expand the continuity equation (\ref{continuity}) and
define at each order a quantity that is exactly conserved along
the fluid flow for adiabatic (isentropic) 
perturbations, i.e., if the ideal
fluid is barotropic.

We begin by considering the evolution of covariant {\em
first-order} perturbations. By applying the spatial derivative
$\cov_{\cv_\perp} $ to Eq.~(\ref{continuity}) we find
\beq
\cov_{\cv_\perp} \dot \rho + 3 (\rho +p)\cov_{\cv_\perp} \dot
\alpha + 3 \dot \alpha (\cov_{\cv_\perp} \rho + \cov_{\cv_\perp}
p)=0. \label{D_continuity}
\eeq
Now we want to invert the time derivative with the {\em space}
derivative along $\cv^a_\perp$. However, before doing so we note
that $\cv^a_\perp$ does not, in general, commute with $u^a$ but one
has
\beq [u,\cv_\perp]^a = -\dot \cv_\parallel u^a
\label{commutator_bis} \eeq
which implies that
\beq
\cov_{\cv_\perp} \dot f = \Tdot{( \cov_{\cv_\perp} f)} +\dot
\cv_\parallel \dot f.
\eeq
Using this relation and Eq.~(\ref{continuity}), we can rewrite
Eq.~(\ref{D_continuity}) as
\beq \Tdot{\left( \cov_{\cv_\perp} \alpha - \frac{
\dot \alpha }{\dot \rho}\cov_{\cv_\perp}\rho \right)} = \frac{3 \dot
\alpha^2}{\dot \rho} \left( \cov_{\cv_\perp} p - \frac{ \dot p}{\dot
\rho}\cov_{\cv_\perp} \rho \right),
\label{zeta_conservation_1}
\eeq
which has the same form as Eq.~(\ref{zeta_a_conservation}) but is
written in terms of scalar quantities instead of covectors.
Indeed, this equation can be also found by projecting
Eq.~(\ref{zeta_a_conservation}) along $e^a$, and making use of
Eqs.~(\ref{commutation_rel}) and (\ref{rel_1}).

It is convenient to re-express Eq.~(\ref{zeta_conservation_1}) by
defining the vector field
\beq
\lv^a \equiv \cv_\perp^a - \frac{ \cov_{\cv_\perp}\rho}{\dot \rho}
u^a=\cv^a - \frac{ \cov_{\cv}\rho}{\dot \rho} u^a \label{lv_def}
\eeq
(the second equality follows from Eq.~(\ref{connecting})), which
lies on a uniform density hypersurface, as one can check by
contracting its definition with $\nabla_a \rho$. Using this
vector, Eq.~(\ref{zeta_conservation_1}) can be rewritten as
\beq
\Tdot{\left( \cov_{\lv} \alpha \right)} = \frac{3 \dot \alpha^2}{\dot
\rho} \cov_{\lv} p.
\label{zeta_conservation_2}
\eeq
Note that although in Eq.~(\ref{zeta_conservation_1}) we have used
covariant derivatives along the {\em spatially projected} vector
$\cv_\perp^a$ to define the perturbations,
because of the second equality of Eq.~(\ref{lv_def}) one can
replace these by covariant derivatives along $\cv^a$.

The quantity
\beq
\cov_\lv \alpha = \cov_{\cv} \alpha - \frac{ \cov_{\cv}\rho}{\dot
\rho} \dot \alpha, \qquad (\cov_\lv \rho=0), \label{rhozero}
\eeq
describes the change of $\alpha$ when going from a worldline to a
neighboring one along the integral curve of $\cv^a$ projected on
uniform density hypersurfaces. This can be seen more clearly by
rewriting the previous expression in terms of the coordinates
$\{\tau,\lambda\}$, which yields
\beq
\label{defzetancrd2} \cov_\lv \alpha=\frac{d
\alpha}{d\lambda}-\frac{\dot \alpha}{\dot{\rho}} \frac{d \rho}{d
\lambda}= \left.
\frac{\partial\alpha}{\partial\lambda}\right|_{\rho},
\eeq
where the second equality follows from the change of variables
$\{\tau,\lambda\} \to \{\rho,\lambda\}$. Thus, $\cov_\lv \alpha$
in Eq.~(\ref{zeta_conservation_2}) is the covariant first-order
perturbation of the integrated expansion $\alpha$ on uniform
density hypersurfaces and $\cov_\lv p$ is the covariant
first-order non-adiabatic pressure perturbation.

By introducing the notation
\beq
\label{zeta1dot} \zeta_{(1)}\equiv \cov_\lv \alpha,
\eeq
and
\beq \Gamma_{(1)}\equiv \cov_\lv p,
\eeq
the evolution equation (\ref{zeta_conservation_2}) can be written as
\beq
\label{z1}\dot{\zeta}_{(1)}=\frac{3\dot{\alpha}^2}{\dot{\rho}}\Gamma_{(1)},
\eeq
which shows that $\zeta_{(1)}$ is conserved if $\Gamma_{(1)}=0$.
This equation is exact and fully nonlinear and generalizes the
large-scale first-order conservation equation of the curvature
perturbation on uniform density hypersurfaces $\zeta$.

From Eq.~(\ref{defzetancrd2}) we can interpret the conservation of
$\zeta_{(1)}$ for a barotropic fluid as follows (see also the
discussion of \cite{Lyth:2003im,proof}). The continuity equation
of an ideal fluid, Eq.~(\ref{continuity}), can be integrated {\em
along} each comoving worldline yielding
\beq
\alpha = - \int \frac{d\rho}{3(\rho+p)}, \qquad (\lambda = {\rm
const}). \label{integration}
\eeq
If the fluid is barotropic, $p=p(\rho)$ is the same function for
all worldlines, and Eq.~(\ref{integration}) can be explicitly
integrated yielding
\beq
\alpha= \tilde \alpha(\rho) + c(\lambda), \label{alpha_baro}
\eeq
where $\tilde \alpha(\rho)$ is a function of $\rho$ and
$c(\lambda)$ is an integration constant (along a worldline) that
may change from a worldline to another, and reflects that $\alpha$ is defined
up to a constant. Since $\cov_\lv \tilde
\alpha=0$, from Eqs.~(\ref{zeta1dot}) and (\ref{defzetancrd2})
this expression yields
\beq
\zeta_{(1)} = \left.\frac{ \partial c}{\partial \lambda}
\right|_\rho,
\eeq
which is constant along the fluid flow.
However, if the fluid is non-barotropic, the equation of state $p=p(\rho,\lambda)$
may vary from a worldline to another implying that the dependence of
$\alpha$ on $\rho$ also changes and $\zeta_{(1)}$, as defined in
Eq.~(\ref{defzetancrd2}), is no longer conserved.

Before extending our analysis to an arbitrary order, we first
consider covariant {\em second-order} perturbations for which we
know the coordinate-based perturbative analogue. As explained
above, in our covariant approach the operator $\cov_{\cv_\perp}$
is used to define perturbations and $\cov_{\lv}$ combines these to
construct perturbations on uniform density hypersurfaces. In order
to find the second-order perturbed evolution equation we can apply
once more $\cov_{\cv_\perp}$ to Eq.~(\ref{D_continuity}). However,
it is more convenient to expand the continuity equation in
perturbations {\em on} uniform density hypersurfaces, applying
directly $\cov_{\lv}$ to Eq.~(\ref{zeta_conservation_2}). Using
the commutation relation
\beq
\label{relation_X2} [\lv,u]^{a}=\Tdot{\left(\frac{\nabla_\cv
\rho}{{\dot{\rho}}}\right)}u^{a},
\eeq
which follows from the definition of $\lv^a$ and implies
\beq
\cov_\lv \dot f = \Tdot{ ( \cov_\lv f)}  +  \Tdot{\left(
\cv_\parallel +
\frac{\cov_{\cv_\perp} \rho}{\dot \rho}\right)}
\dot f, \label{relation_X}
\eeq
we find a second-order evolution equation,
\beq
\dot \zeta_{(2)} =\frac{3 \dot \alpha^2}{\dot \rho}
\Gamma_{(2)} + \frac{6 \dot \alpha}{\dot \rho} \dot \zeta_{(1)}
\Gamma_{(1)},\label{zeta2dot}
\eeq
where we have defined the  covariant second-order perturbation of
$\alpha$ on uniform density hypersurfaces as
\beq
\label{zeta2} \zeta_{(2)}\equiv\cov_\lv^2\alpha,
\eeq
and the covariant second-order non-adiabatic pressure perturbation
as
\beq
\label{gamma2} \Gamma_{(2)}\equiv \cov_\lv^2p.
\eeq
Expanding the definition (\ref{zeta2}) in terms of spatial
perturbations one obtains
\beq
\zeta_{(2)}=\cov_{\cv_\perp}^2
\alpha-\frac{\dot{\alpha}}{\dot{\rho}}\cov_{\cv_\perp}^2\rho-
\frac{2}{\dot{\rho}}\cov_{\cv_\perp} \rho \left[
\Tdot{(\cov_{\cv_\perp} \alpha)}-\frac{\dot{\alpha}}{\dot{\rho}}
\Tdot{(\cov_{\cv_\perp}\rho)}\right]+\frac{1}{\dot{\rho}}
\Tdot{\left(\frac{\dot{\alpha}}{\dot{\rho}}\right)}
(\cov_{\cv_\perp}\rho)^2,\label{zeta2_bis}
\eeq
and the explicit expansion of $\Gamma_{(2)}$ can be read from
Eq.~(\ref{zeta2_bis}) by replacing $\alpha$ by $p$.

Equation (\ref{zeta2dot}) is the evolution equation for
$\zeta_{(2)}$. It implies that $\zeta_{(2)}$ is conserved on all
scales if the first and second-order non-adiabatic pressure
perturbations vanish, i.e., $\Gamma_{(1)}=\Gamma_{(2)}=0$. The
form of Eq.~(\ref{zeta2_bis}) and the evolution equation
(\ref{zeta2dot}) mimic and generalize the large-scale result of
the second-order coordinate approach \cite{Malik:2003mv}.

We are now ready to extend our analysis to arbitrary order by
defining the covariant $n$-th order perturbation of $\alpha$ on
uniform density hypersurfaces as
\beq
\label{defzetan}\zeta_{(n)}\equiv \cov_\lv^n\alpha = \left(
\cov_\cv -\frac{\cov_\cv \rho}{\dot \rho} \cov_u \right)^n \alpha,
\eeq
and the $n$-th order non-adiabatic pressure perturbation as
 \beq
\label{defgamman} \Gamma_{(n)}\equiv \cov_\lv^n p. \eeq To find
the evolution equation of $\zeta_{(n)}$, one can apply the
operator $\cov_\lv^n$ to the continuity equation
(\ref{continuity}) and recursively use the commutation relation
(\ref{relation_X}) and $\cov_\lv \rho=0$ to invert the time and
space derivatives. After a series of straightforward manipulations
one obtains \beq \label{zetandot}
\dot{\zeta}_{(n)}=\frac{3}{\dot{\rho}}
\sum_{l=0}^{n-1}\sum_{m=0}^{n-l-1} \frac{(n-1)!}{l!m!(n-l-m-1)!}
\dot{\zeta}_{(n-l-m-1)}\dot{\zeta}_{(m)}\Gamma_{(l+1)}, \eeq where
we have defined \beq \zeta_{(0)} \equiv \alpha. \eeq For any order
of the perturbative expansion defined in
Sec.~\ref{sec:definition}, this evolution equation shows that
$\zeta_{(n)}$ is conserved on all scales if the fluctuations are
adiabatic up to this order, i.e., $\Gamma_{(k)}=0,~k=1,\ldots,n$.
Indeed, for a barotropic fluid, Eq.~(\ref{alpha_baro}) implies
that $\zeta_{(n)}$ is conserved along a worldline at all orders.
The definition of $\zeta_{(n)}$ Eq.~(\ref{defzetan}) and its
evolution equation (\ref{zetandot}) are among our main results.

\section{Comparison with the coordinate-based approach}
\label{sec:coordinates}

Most of the studies of inhomogeneities in cosmology have been done
in the coordinate-based perturbation theory. Thus, it is important
to establish a connection between our covariant perturbations and
the quantities used in the coordinate approach.

In this section we construct at arbitrary order in the
coordinate-based perturbation theory the expression and evolution
equation of the perturbation of the integrated expansion on
uniform density hypersurfaces, which on large scales coincides
with the curvature perturbation on uniform density hypersurfaces,
thus extending the previously known first and second-order
results. Then, we explicitly expand the covariant variable
$\zeta_{(n)}$ given in Eq.~(\ref{defzetan}) in terms of perturbations in a coordinate system and
show that, in the uniform energy density gauge, it reduces on
large scales to the spatial gradient of the $n$-th order uniform
density curvature perturbation.

\subsection{Perturbation theory and gauge transformations}
\label{sec:gauge}
\def\y{y}

In the standard coordinate-based perturbation theory (see e.g.
\cite{Nakamura,Bruni}) one considers a $5$-dimensional manifold
$\mc{N}=\mc{M}\times\mathbb{R}$, each $\mc{M}$ being labelled by the
continuous parameter $\y$. Each submanifold $\mc{M}_\y$, together
with the tensor fields $T_{\y}$ living on it, describes a
spacetime model which interpolates between an ideal FLRW
background, at $\y=0$, and the real inhomogeneous universe, at
$\y=1$. The real universe can then be described approximately by
an expansion in the parameter $\y$ around the background solution.
In the following we choose the parameter $\y$ to be the fifth
coordinate on $\mc{N}$, $x^4=\y$, and use capital indices $A,B$
running from $0$ to $4$ to denote the components of a tensor field
on $\mc{N}$.

To define the perturbation of a tensor field $T$ around the
background spacetime $\y=0$, one needs a map between the
submanifolds $\mc{M}_{\y}$, which can be constructed as the flow
${\cal X}_{\y}:\mc{M}_{0}\rightarrow\mc{M}_{\y}$ of a vector field
$X^{A}$ defined on $\cal N$ such that $X^4=1$ everywhere. Thus,
$X^A$ is always transverse to ${\cal M}_\y$ and connects different
leaves of the foliation of $\cal N$. The perturbation of a tensor
field $T$ can then be defined as \cite{Bruni}
\beq
\label{perturbation} (\Delta_X T)_y \equiv  {\cal X}^{*}_{\y}
\left. T \right|_{0}- \left. T \right|_0,
\eeq
where the subscript $0$ denotes the restriction to the background
spacetime $\mc{M}_0$ and we recall that ${\cal X}^{*}_{\y}T$ is
the pull-back of $T_\y$. This can  be  taylor-expanded in
the parameter $\y$ as
\beq
\label{expansion}(\Delta_X T)_{\y}=\sum_{n=1}^{\infty}
\frac{\y^n}{n!}\left. \lie_X^n T\right|_{0},
\eeq
and the $n$-th order term of this expansion defines the $n$-th
order perturbation of $T$,
\beq
\label{deltan}\delta_{ X}^n T \equiv  \lie_X^n \left.
T\right|_{0}.
\eeq

The definitions of the perturbation of $T$ and of its perturbative
expansion depend on the choice of the vector field $X^{A}$, or
equivalently of the diffeomorphism ${\cal X}_{\y}$. This is
commonly referred to as the choice of {\em gauge} and therefore
Eq.~(\ref{deltan}) defines the $n$-th order perturbation of $T$ in
the gauge $X^{A}$. Instead of $X^{A}$, one can define the
perturbation of $T$ by using another vector field $Y^{A}$ with
$Y^4=1$,
\beq
\label{perturbation_2} (\Delta_Y T)_{\y}\equiv {\cal Y}^{*}_{\y}
\left. T \right|_{0}- \left. T \right|_{0},
\eeq
where ${\cal Y}_{\y}$ is the flow generated by $Y^A$. One can
expand this equation similarly to Eq.~(\ref{expansion}) and define
the $n$-th order perturbation of $T$ in the gauge $Y^{A}$ as
\beq
\label{deltanY} \delta_Y^n T \equiv  \left. \lie_Y^n
T\right|_{0}.
\eeq

The transformation from the gauge $X^{A}$ to $Y^{A}$ is generated
by the diffeomorphism $({\cal Y}_{-\y} \circ {\cal
X}_\y):\mc{M}_0\to\mc{M}_0$. One can taylor-expand its action on
$(\Delta_y T)_X$ and explicitly work out how the perturbations
transform order by order, defining at each order a {\em generator}
of the gauge transformation as a vector field living on the
background spacetime \cite{Bruni}. For our purposes, it is more
convenient to express the gauge transformation in terms of the
vector field $\xi^{A}$ defined as
\beq
\label{xi} \xi^{A} \equiv  Y^{A}- X^{A},
\eeq
which will be called here the {\em total generator} of the
gauge-transformation to distinguish it from the generators defined
order by order. Equation (\ref{xi}) defines $\xi^{A}$ for each
value of $\y$ and can therefore be used to generate gauge
transformations at arbitrary order. Note that $\xi^4$ vanishes
identically due to the choice $X^{4}=Y^{4}=1$ showing that $\xi^A$
evaluated for a given $\y$ is always tangent to ${\cal M}_\y$.

The $n$-th order perturbation of a tensor field $T$ in the gauge
$Y^{A}$ can now be written in terms of the perturbations in the
gauge $X^{A}$ and the total generator $\xi^{A}$ as
\beq
\label{compact_n} \delta_Y^n T=\left. \lie^n_Y
T\right|_0 = \left.(\lie_X + \lie_{\xi})^nT \right|_{0}.
\eeq
One can employ this compact formula to
derive the gauge-invariant expression of the perturbation of $T$
in the gauge $Y^{A}$ at any order.

The gauge transformations derived from Eq.~(\ref{compact_n}) are
equivalent to those derived in \cite{Bruni}. When expanding the
right hand side of Eq.~(\ref{compact_n}) to $n$-th order, one
finds combinations of commutators of the vector fields $X^A$ and
$\xi^A$ that are equivalent to the $n$ independent generators
defined in \cite{Bruni}. Indeed, this can be demonstrated up to
third order by explicitly expanding Eq.~(\ref{compact_n}). By
employing the following useful expression for the commutator of
two Lie derivatives along the vector fields $v^A$ and $w^A$, \beq
[\lie_v, \lie_w] = \lie_{[v,w]}, \label{lie_commutator}
\eeq
this yields
\baq \delta_Y T
&=& \delta_X T + \lie_{\xi_{(1)}} \left. T\right|_0, \label{gt_1} \\
\delta_Y^2 T 
&=&  \delta_X^2 T + ( \lie_{\xi_{(2)}}+\lie^2_{\xi_{(1)}}) \left. T\right|_0
+2\lie_{\xi_{(1)}} \delta_X T, \label{gt_2} \\
\delta_Y^3 T &=&  \delta_X^3 T + ( \lie_{\xi_{(3)}}+3
\lie_{\xi_{(1)}} \lie_{\xi_{(2)}} + \lie_{\xi_{(1)}}^3) \left. T\right|_0 +
3 (\lie_{\xi_{(2)}} + \lie_{\xi_{(1)}}^2 ) \delta_X T + 3
\lie_{\xi_{(1)}} \delta_X^2 T, \cr &&\label{gt_3} \eaq
where we have defined the first three generators of the gauge
transformations as
\baq \xi^A_{(1)} &\equiv& \left. \xi^A \right|_0
= \left. Y^A-X^A \right|_0, \label{xi_1} \\
\xi^A_{(2)}&\equiv& \left. [X,\xi]^A \right|_0 =
\left. [X,Y]^A \right|_0, \label{xi_2}\\
\xi^A_{(3)}&\equiv& \left. [X-\xi,[X,\xi]]^A \right|_0 = \left.
[2X-Y,[X,Y]]^A \right|_0,  \label{xi_3} \eaq
and we have used Eq.~(\ref{xi}) to rewrite them in the second
equalities in the familiar form in terms of $X^{A}$ and $Y^{A}$,
given in \cite{Bruni}.

\subsection{Evolution of perturbations}
\label{sec:uniform_integrated}

In this section we derive the gauge-invariant expression and the evolution
equation of the perturbation of the integrated expansion (or number of
e-folds) on uniform density hypersurfaces at arbitrary order using
Eq.~(\ref{compact_n}). We focus on this quantity, instead of the
curvature perturbation on uniform density hypersurfaces, because
it is the conserved quantity that naturally arises when expanding
the continuity equation. However, on large scales, i.e., neglecting spatial
gradients as well as vector and tensor perturbations, as in
the so-called separate universe approach
\cite{Sasaki:1998ug,Rigopoulos:2003ak,proof,Langlois:2005ii},
the
curvature perturbation coincides with the perturbation in the
integrated expansion. Therefore, on these scales the uniform
density integrated expansion coincides with the uniform density
curvature perturbation, allowing us to establish also for the
latter an expression on large scales at arbitrary order.

In the following we choose coordinates on the background manifold
$\mc{M}_0$ such that the FLRW metric reads
\beq
ds^2=-dt^2+a^2(t) \gamma_{ij} dx^i dx^j,
\eeq
where $\gamma_{ij}$ is the homogeneous spatial metric and the
four-velocity of the fluid is
\beq
\bar u^{a} = (1,0,0,0).
\eeq

The uniform energy density gauge is defined by requiring that the
perturbations of the energy density vanish to all orders. This
determines the temporal gauge, or time-slicing, and additional
conditions are needed to fix the remaining three degrees of freedom
that correspond to the spatial gauge or threading. In the following
we denote by $Y^{A}$ the gauge with uniform energy density slicing
and comoving threading and derive the gauge-invariant expression for
the perturbations of $\alpha$ in this gauge.

The conditions defining the gauge $Y^{A}$ can be expressed as
\baq
\label{unienergy} \lie_Y\rho&=&0,\\
\label{comoving} \lie_Yu^{A}&\propto&u^{A}. \eaq The first
condition guarantees that the perturbations of the energy density
vanish to all orders,
\beq
\label{order} \delta_Y^n \rho=0,
\eeq
and the second condition sets the perturbations of the spatial
components of the fluid four-velocity to zero
\beq
\delta^n_Yu^{i}=0.
\eeq

By substituting Eq.~(\ref{xi}) into Eqs.~(\ref{unienergy}) and
(\ref{comoving}), we derive a set of conditions on the total
generator $\xi^{A}$ that define the gauge transformation between
the gauge $Y^{A}$ and a generic gauge $X^A$. To simplify the
analysis, we decompose $\xi^{A}$ into parts orthogonal and
parallel to the four-velocity $u^{A}$ as
\beq
\label{dec_xi} \xi^{A}=\xi_{\parallel}u^{A}+\xi_{\perp}^{A},\qquad
(\xi_{\perp}^{A} = h^{A}_{\ B}\xi^{B},~\xi_{\parallel} =
-u^{A}\xi_{A}).
\eeq
Using the condition for uniform energy density slicing,
Eq.~(\ref{unienergy}), we can rewrite $\xi^A$ as
\beq
\label{xiuni}
\xi^A=-\frac{\lie_X\rho}{\dot{\rho}}u^{A}+\tilde{\xi}^{A},
\eeq
where we have defined
\beq
\tilde{\xi}^{A}\equiv\xi^{A}_{\perp}-\frac{\lie_{\xi_{\perp}}\rho}{\dot{\rho}}u^{A}.
\label{tilde_xi}
\eeq
Furthermore, the condition for comoving threading,
Eq.~(\ref{comoving}), yields a condition on the Lie derivative of
$\tilde \xi^A$ along $u^A$, i.e.,
\beq
\lie_u \tilde \xi^A - \lie_X u^A  \propto u^A.
\label{comoving_condition}
\eeq

The expression for $\xi^{A}$ (\ref{xiuni}) with the condition
(\ref{comoving_condition}) involves only perturbations in the
gauge $X^{A}$ and by substituting it into Eq.~(\ref{compact_n}) we
obtain the gauge-invariant expression for $\delta_Y^n\alpha$,
\beq
\delta^n \zeta \equiv \delta_Y^n
\alpha=\left.\left(\lie_X-\frac{\lie_X\rho}{\dot{\rho}}\lie_u+\lie_{\tilde{\xi}}\right)^n
\alpha\right|_0, \label{gizeta}
\eeq
where we have defined $\delta^n \zeta$ as the gauge-invariant $n$-th order
perturbation of the integrated expansion on uniform density
hypersurfaces.

Using Eq.~(\ref{gizeta}) one can straightforwardly rederive the
familiar first and second-order expressions for $\delta \zeta$ and
$\delta^2 \zeta$ and even go to higher orders. At first order
$n=1$, using the definition of perturbations (\ref{deltan}),
Eq.~(\ref{gizeta}) yields
\beq
\delta \zeta=\delta_X\alpha-\frac{H}{\bar \rho'}\delta_X\rho,
\label{delzeta1}
\eeq
where $\bar \rho\equiv \left.\rho\right|_0$ is the background
value of the energy density, $H\equiv a'/a= \bar \alpha'$ the
background Hubble parameter and the prime ${}'$ denotes the
derivative with respect to the time coordinate $t$. The vector
$\tilde \xi^{A}$ specifying the threading does not appear in this
expression because $\left.\lie_{\tilde{\xi}} \alpha\right|_0=0$.
On large scales, $\delta_X \alpha$ is equivalent to the
first-order curvature perturbation and Eq.~(\ref{delzeta1})
coincides with the well-known gauge-invariant expression for the
curvature perturbation on uniform density hypersurfaces.

At second order,  $n=2$, by using Eq.~(\ref{lie_commutator}) for
the commutator of two Lie derivatives, Eq.~(\ref{gizeta}) becomes
\beq
\label{gizeta2} \delta^2 \zeta= \left. \left[
\left(\lie_X-\frac{\lie_X\rho}{\dot{\rho}}\lie_u
\right)^2\alpha+2\lie_{\tilde{\xi}}\left(\lie_X
\alpha-\frac{\lie_X\rho}{\dot{\rho}}\dot \alpha \right)+
\lie_{[X-({\lie_X\rho}/{\dot{\rho}})u,\tilde{\xi}]}\alpha+\lie_{\tilde{\xi}}^2\alpha
\right] \right|_0.
\eeq
The last two terms vanish on the background manifold as one can
show by using $\lie_{\tilde \xi} \rho=0$ and $ \tilde{\xi}^0
\left. \right|_0=0$, while the second term is simply
$2\lie_{\tilde \xi_{(1)}} \delta \zeta$. Furthermore, when
commuting $\lie_u$ and $\lie_X$ in the first term on the right
hand side of Eq.~(\ref{gizeta2}) using Eq.~(\ref{lie_commutator}),
one encounters terms proportional to the commutator $[X,u]^{A}$.
These terms do not vanish but appear in such a combination
that they cancel when evaluated on the background. After these
manipulations one arrives at
\beq
\label{delzeta2} \delta^2 \zeta = \delta_X^2 \alpha -
\frac{H}{\bar\rho'} \delta_X^2 \rho - \frac{2}{\bar\rho'}\delta_X
\rho \left(\delta_X \alpha' - \frac{H}{\bar\rho'} \delta_X \rho'
\right)+\frac{1}{\bar\rho'} \left(\frac{H}{\bar\rho'} \right)'
\delta_X \rho^2 + 2{\tilde \xi}_{(1)}^{i}\partial_i\delta \zeta,
\eeq
where
\beq
\label{1st_gen} {\tilde \xi}_{(1)}^{i} \equiv \left. {\tilde \xi}^{i}
\right|_0
\eeq
is the generator of the first-order spatial gauge transformation
from the gauge $X^{A}-(\lie_X\rho/{\dot{\rho}})u^{A}$ to $Y^{A}$
and for convenience we adopt the notation $\delta_X \rho' \equiv
(\delta_X \rho)'$.

The first-order generator ${\tilde \xi}_{(1)}^{i}$ can be
expressed in terms of the perturbations of $u^{A}$ restricting the
spatial components of Eq.~(\ref{comoving_condition}) to the
background manifold, which yields
\beq
{\tilde \xi}_{(1)}^{i}{}' = \delta_X u^i. \label{xi_i_v}
\eeq

Without the last term on the right hand side, Eq.~(\ref{delzeta2})
is equivalent to the second-order perturbation of the integrated
expansion on uniform density hypersurfaces defined in
\cite{Langlois:2005ii}. Here, in addition we were
able to account also for the last term of Eq.~(\ref{delzeta2})
that was present in \cite{Malik:2003mv}, coming from the choice of
threading. On large scales $\delta^2 \zeta$ coincides with the
curvature perturbation on uniform density hypersurfaces defined by
Malik and Wands. (More precisely, it coincides with
$\zeta^{(2)}_{\rm MW} -\zeta^{(1)}_{\rm MW}{}^2$.)

One of the advantages of the compact expression (\ref{gizeta}) is
that one can straightforwardly work out the explicit
gauge-invariant expression for $\delta^n \zeta$ at arbitrary
order. To demonstrate this we give the explicit gauge-invariant
expression of the {\em third order} perturbations of the
integrated expansion on uniform density hypersurfaces. This can be
computed similarly to the second-order case and one finds
\baq\label{thirdorder}
\delta^3
\zeta&=&\delta_X^3\alpha-\frac{H}{\bar{\rho}'}\delta_X^3\rho-
\frac{3}{\bar{\rho}'}\left({\delta_X\alpha}'-\frac{H}{\bar{\rho}'}{\delta_X\rho}'\right)\left(\delta_X^2\rho
-2\frac{\delta_X\rho{\delta_X\rho}'}
{\bar{\rho}'}+\frac{\bar{\rho}''}{\bar{\rho}'{}^2}{\delta_X\rho}^2\right)\nonumber
\\&&-3\frac{\delta_X\rho}{\bar{\rho}'}
\left(\delta_X^2\alpha'-\frac{H}{\bar{\rho}'}{\delta_X^2\rho}'\right)+3\frac{\delta_X\rho^2}{\bar{\rho}'{}^2}
\left(\delta_X\alpha''-\frac{H}{\bar{\rho}'}\delta_X\rho''\right)\nonumber
\\&&+\frac{3}{\bar{\rho}'}\left(\frac{H}{\bar{\rho}'}\right)'\delta_X^2\rho \delta_X\rho
-\frac{6}{\bar{\rho}'{}^2}\left(\frac{H}{\bar{\rho}'}\right)'\delta_X\rho{}^2
\delta_X\rho'-
\frac{1}{\bar{\rho}'}\left[\frac{1}{\bar{\rho}'}\left(\frac{H}{\bar{\rho}'}\right)'\right]'\delta_X\rho^3\nonumber
\\&&+3\left[\tilde \xi_{(1)}^{i}\partial_{i}\delta^2 \zeta - \tilde \xi^{i}_{(1)}\partial_{i} \left(\tilde
\xi^{j}_{(1)}\partial_{j} \delta\zeta \right)+\tilde
\xi^{i}_{(2)}{\partial_{i}}\delta\zeta
 \right] , \eaq
where
\beq
\tilde \xi_{(2)}^{i}\equiv \left.
\left[X-\frac{\lie_X\rho}{\dot{\rho}}u,\tilde{\xi}\right]^{i}\right|_0
\eeq
is the generator of the second-order spatial gauge transformation.
The generator $\tilde \xi_{(2)}^{i}$ can be written in terms of
the perturbations of the four-velocity and energy density by
taking the Lie derivative with respect to $X^A$ of
Eq.~(\ref{comoving_condition}) and restricting the resulting
equation on the background manifold. After some manipulations this
yields
\beq
\tilde \xi_{(2)}^{i}{}'= \delta_X^2 u^i -2 \delta_X u^0 \delta_X u^i
+ \left(\int dt \delta_X u^j  \right) \partial_j \delta_X u^i -
\delta_X u^j
\partial_j \left(\int dt \delta_X u^i  \right) - \left(\frac{\delta_X \rho}{\bar \rho'} \delta_X u^i\right)',
\eeq
where we have used Eq.~(\ref{xi_i_v}) to replace the first-order
generator $\tilde \xi_{(1)}^i$.

In order to derive the evolution equation of $\delta^n\zeta$ one
can perturb the continuity equation (\ref{continuity}) in the
gauge $Y^{A}$. By virtue of Eq.~(\ref{comoving}), the derivation
is formally analogous to the derivation of the covariant evolution
equation of $\zeta_{(n)}$, Eq.~(\ref{zetandot}), which was
obtained in Sec.~\ref{sec:evolution}  by acting on the continuity
equation with $\cov^n_{\lv}$ and using the commutation relation
(\ref{relation_X2}). Thus, one obtains
\beq
\label{pzetandot} \delta^n\zeta'=\frac{3}{\bar{\rho}'}
\sum_{l=0}^{n-1}\sum_{m=0}^{n-l-1}\frac{(n-1)!}{l!m!(n-l-m-1)!}
\delta^{n-l-m-1}\zeta'\delta^m\zeta'\delta^{l+1}\Gamma,
\eeq
where
\beq
\delta^0 \zeta' \equiv H,
\eeq
and the gauge-invariant $n$-th order non-adiabatic pressure
perturbation is defined as
\beq
\delta^n\Gamma\equiv\delta^n_Yp.
\eeq
As expected, the evolution equation of $\delta^n\zeta$ is exactly
of the same form as that for $\zeta_{(n)}$, Eq.~(\ref{zetandot}).

On large scales where the perturbation of the integrated expansion
and the curvature perturbation coincide to all orders,
$\delta^n\zeta$ is equivalent to the $n$-th order curvature
perturbation on uniform density hypersurfaces. In particular,
Eq.~(\ref{pzetandot}) shows that, on large scales, for adiabatic
perturbations, the curvature perturbation on uniform density
hypersurfaces is approximatively conserved at all orders.

By choosing in the gauge $X^A$ a threading such that $\tilde \xi^A
=0$, the expression (\ref{gizeta}) for $\delta^n \zeta$ has
exactly the same form as the covariant definition of
$\zeta_{(n)}$, Eq.~(\ref{defzetan}), once the perturbations
$\delta_X^n\alpha$ and $\delta_X^n\rho$ are replaced by the
covariant perturbations $\cov_{\cv_\perp}^n\alpha$ and
$\cov_{\cv_\perp}^n\rho$. Furthermore, since $\zeta_{(n)}$ and
$\delta^n\zeta$ satisfy the same evolution equation, we conclude
that $\zeta_{(n)}$ provides the {\em covariant}
generalization of the coordinate-based
perturbative quantity $\delta^n \zeta$.

\subsection{Coordinate-based expansion of the covariant perturbations}

Covariant quantities can be expanded in terms of perturbations in
a coordinate system. In the covariant approach of Ellis and Bruni,
this expansion has been done in \cite{Bruni:1992dg}. However, it
was restricted to first order in the coor\-di\-na\-te-based
perturbations, where covariant quantities are automatically
gauge-invariant. At higher order, covariant quantities are not
necessarily gauge invariant (see also the discussions in
\cite{Langlois:2005ii} and \cite{Langlois:2006vv}).

In the following we expand our covariant variable
$\zeta_{(n)}$ in the coor\-di\-na\-te-based perturbations. We
first do this in the uniform density slicing and comoving threading $Y^A$, defined in
Sec.~\ref{sec:uniform_integrated}, and then consider a general gauge
 $X^A$. In order to simplify our
analysis, we choose the gauge $Y^A$ such that it commutes with
$\lv^A$, i.e.,
\beq
\label{lvgauge} \lie_Y \lv^{A}=0,
\eeq
which implies that $\lv^a$ is unperturbed to all orders in the
gauge $Y^A$,
\beq
\label{lv_unpert}\delta_Y^n \lv^a = 0.
\eeq
This condition is compatible with Eqs.~(\ref{unienergy}) and
(\ref{comoving}) and it is conserved during the time evolution.

With this assumption, one can perturb the definition of
$\zeta_{(n)}$, Eq.~(\ref{defzetan}), to the $m$-th order using
Eq.~(\ref{deltanY}) and commuting $Y^A$ with $\tilde e^A$ using
the condition (\ref{lvgauge}) yields
\beq
\label{delzeta} {\delta_Y^m \zeta_{(n)}} =\lie_\lv^n \delta^m
\zeta.
\eeq
Furthermore, one can use Eq.~(\ref{lv_unpert}) to express the Lie
derivative along $\lv^a$ in terms of $\bar e^i$, the background
component of $e^a$, which can be shown to be constant in time,
$\bar e^i{}'=0$, due to Eq.~(\ref{commutation_rel}). Finally, one
obtains
\beq
\label{rel_Y_X}{\delta_Y^m \zeta_{(n)}} =(\bar e^i
\partial_i)^n \delta^m \zeta.
\eeq
As expected, $\delta_Y^m \zeta_{(n)}$  reduces to $n$-th order gradients of
$\delta^m\zeta$ projected along $\bar e^i$. On the left hand side
of this equation, the $m$-th order perturbation of $\zeta_{(n)}$
on uniform density hypersurfaces and comoving threading can be written in
gauge-invariant form using Eq.~(\ref{compact_n}), and on the right hand side
$\delta^m\zeta$ is also gauge-invariant.

Now we want to consider the expansion of $\delta^m\zeta_{(n)}$ in
a {\em generic gauge} $X^A$. At first order in the perturbations,
using the gauge transformation from $Y^A$ to $X^A$
(\ref{compact_n}) one finds
\beq
\delta_X\zeta_{(n)}= {\delta_Y\zeta_{(n)}}. \label{zeta_n_tilde}
\eeq
Indeed, $\zeta_{(n)}$ has no background value and at first order
$\delta_X \zeta_{(n)}$ is automatically gauge-invariant by
Stewart-Walker Lemma \cite{Stewart:1977}. Thus
$\delta_X\zeta_{(n)}$ is simply related to the gradients of
$\delta \zeta$ by
\beq
{\delta_X \zeta_{(n)}} =(\bar e^i
\partial_i)^n \delta \zeta.
\eeq

However, one does not expect such a simple relation to hold at
higher order perturbations $m>1$. For example, one can consider
the second-order perturbation of $\zeta_{(1)}$ in a generic gauge.
Using Eq.~(\ref{compact_n}) one obtains
\beq
{\delta_Y^2 \zeta_{(1)}} = \delta_X^2 \zeta_{(1)} - \frac{2\delta_X
\rho}{\bar \rho'} \delta_X \zeta_{(1)}{}' + 2\tilde \xi_{(1)}^i
\partial_i \delta_X \zeta_{(1)},
\eeq
where $\tilde \xi_{(1)}^i$ is defined in Eq.~(\ref{1st_gen}) and
explicitly given for the comoving threading in (\ref{xi_i_v}).
Replacing this expression in Eq.~(\ref{rel_Y_X}) for $m=2$ and
$n=1$, and using again (\ref{rel_Y_X}) for $m=1$ and $n=1$ with
(\ref{zeta_n_tilde}) to rewrite $\delta_X \zeta_{(1)}$ in terms of
$\delta \zeta$, yields
\beq
\delta_X^2 \zeta_{(1)} =\bar e^i
\partial_i \delta^2 \zeta + \frac{2\delta_X \rho}{\bar
\rho'} \bar e^i
\partial_i \delta \zeta '- \bar e^j
\partial_j \left(2\tilde
\xi_{(1)}^i \partial_i \delta \zeta \right).
\eeq
In a general gauge, for $m>1$, the perturbation of $\zeta_{(n)}$
do not reduce to gradients of $\delta^n\zeta$ alone but also
include terms proportional to the perturbations in the energy
density and the vector $\tilde \xi^i$.

\section{Conclusion}
\label{sec:conclusion}

In this paper we develop a covariant generalization of the
relativistic cosmological perturbation theory by defining the
perturbation of a scalar quantity as its fluctuation along a curve
connecting two comoving observers in the real inhomogeneous
universe. We also extend the formalism to describe perturbations
of tensor fields. 
These perturbations are fully nonlinear. Being
covariantly defined in the real universe, they have a clear physical
interpretation and are not hampered by gauge subtleties.

We use this covariant formalism to define a scalar variable
$\zeta_{(n)}$, as given in Eq.~(\ref{defzetan}), which is the covariant generalization of the $n$-th order
curvature perturbation on uniform density hypersurfaces defined in
the coordinate approach. The variable $\zeta_{(n)}$ is
covariantly constructed in such a way that it describes the $n$-th order
fluctuation of the integrated expansion (or number of e-folds)
$\alpha$ on uniform density hypersurfaces. By using the
continuity equation we derive the evolution equation of
$\zeta_{(n)}$, given in Eq.~(\ref{zetandot}), at arbitrary order in the
covariant perturbations. We also
show that if the fluctuations are adiabatic, i.e., for an
ideal and barotropic fluid, $\zeta_{(n)}$ is  exactly
conserved on all scales.

To show that $\zeta_{(n)}$ generalizes the $n$-th order uniform
density curvature perturbation, in Sec.~\ref{sec:coordinates} we first
present a compact method to construct gauge-invariant
expressions for $n$-th order perturbations
in the standard perturbation theory.
 We then find the $n$-th order perturbation of
the integrated expansion on uniform density hypersurfaces, denoted as
$\delta^n \zeta$ and given in Eq.~(\ref{gizeta}), which {\em on large scales}
coincides with the
curvature perturbation on uniform density hypersurfaces.  Moreover, we
derive a conservation equation
for $\delta^n \zeta$ which is formally the same as the corresponding equation
for
 $\zeta_{(n)}$. Thus we conclude that the conserved covariant
quantities $\zeta_{(n)}$
are for each $n$ the proper generalizations of the analogous quantities
defined in the standard
coordinate-based approach.

The covariant cosmological perturbation theory developed in the present paper
has several advantages. It allows one to construct nonlinear quantities
mimicking those of the standard coordinate-based perturbation theory and
derive their fully nonlinear evolution equations, without making use of
approximations. Furthermore, it provides a clear insight of the conservation
equations. Moreover, the fact that the perturbations are quantities in the
real universe, makes it easy to connect them to observable quantities.

\begin{acknowledgments}
We thank David Langlois for useful comments.
KE and FV would like to thank the Galileo
Galilei Institute for Theoretical Physics for the hospitality.
JH is partially supported by the Magnus Ehrnrooth Foundation
and SN by the Graduate School in Particle and Nuclear Physics.
KE wishes to acknowledge the Academy of Finland grant 110534.
\end{acknowledgments}


\begin{thebibliography}{99}

\bibitem{Lifshitz:1963ps}
  E.~M.~Lifshitz and I.~M.~Khalatnikov,
  Adv.\ Phys.\  {\bf 12}, 185 (1963).

\bibitem{Bardeen:1980kt}
  J.~M.~Bardeen,
  Phys.\ Rev.\ D {\bf 22}, 1882 (1980).

\bibitem{Kodama:1985bj}
  H.~Kodama and M.~Sasaki,
  Prog.\ Theor.\ Phys.\ Suppl.\  {\bf 78}, 1 (1984).

\bibitem{Mukhanov:1990me}
  V.~F.~Mukhanov, H.~A.~Feldman and R.~H.~Brandenberger,
  Phys.\ Rept.\  {\bf 215}, 203 (1992).

\bibitem{Bruni}
  M.~Bruni, S.~Matarrese, S.~Mollerach and S.~Sonego,
  Class.\ Quant.\ Grav.\  {\bf 14} (1997) 2585
  [arXiv:gr-qc/9609040].

\bibitem{secondorder}
  S.~Matarrese, S.~Mollerach and M.~Bruni,
  Phys.\ Rev.\ D {\bf 58}, 043504 (1998) [arXiv:astro-ph/9707278].


\bibitem{Bartolo}
  N.~Bartolo, E.~Komatsu, S.~Matarrese and A.~Riotto,
  Phys.\ Rept.\  {\bf 402}, 103 (2004)
  [arXiv:astro-ph/0406398].

\bibitem{Noh:2004bc}
  H.~Noh and J.~Hwang,
  Phys.\ Rev.\ D {\bf 69}, 104011 (2004)
  [arXiv:astro-ph/0305123].

\bibitem{Nakamura}
  K.~Nakamura,
  arXiv:gr-qc/0605107; {\em ibid.},
  arXiv:gr-qc/0605108.


\bibitem{Salopek:1990jq}
  D.~S.~Salopek and J.~R.~Bond,
  Phys.\ Rev.\ D {\bf 42} (1990) 3936.



\bibitem{Comer:1994np}
  G.~L.~Comer, N.~Deruelle, D.~Langlois and J.~Parry,
  Phys.\ Rev.\ D {\bf 49}, 2759 (1994).

\bibitem{Sasaki:1998ug}
  M.~Sasaki and E.~D.~Stewart,
  Prog.\ Theor.\ Phys.\  {\bf 95}, 71 (1996)
  [arXiv:astro-ph/9507001];
  M.~Sasaki and T.~Tanaka,
  Prog.\ Theor.\ Phys.\  {\bf 99}, 763 (1998)
  [arXiv:gr-qc/9801017];
  A.~A.~Starobinsky, Phys.\ Lett.\ B {\bf 117}, 175 (1982);
  {\em ibid.}
 JETP Lett.\ {\bf 42}, 152 (1985) [Pisma Zh.\ Eksp.\ Teor.\ Fiz.\ {\bf
 42}, 124 (1985)].


\bibitem{Lyth:2003im}
  D.~H.~Lyth and D.~Wands,
  Phys.\ Rev.\ D {\bf 68}, 103515 (2003)
  [arXiv:astro-ph/0306498].

\bibitem{Rigopoulos:2003ak}
  G.~I.~Rigopoulos and E.~P.~S.~Shellard,
  Phys.\ Rev.\ D {\bf 68}, 123518 (2003)
  [arXiv:astro-ph/0306620].

\bibitem{proof}
  D.~H.~Lyth, K.~A.~Malik and M.~Sasaki,
  JCAP {\bf 0505} (2005) 004
  [arXiv:astro-ph/0411220].




\bibitem{Ellis:1989jt}
  G.~F.~R.~Ellis and M.~Bruni,
  Phys.\ Rev.\ D {\bf 40}, 1804 (1989).

\bibitem{Bruni:1992dg}
  M.~Bruni, P.~K.~S.~Dunsby and G.~F.~R.~Ellis,
  Astrophys.\ J.\  {\bf 395}, 34 (1992).



\bibitem{Langlois:2005ii}
  D.~Langlois and F.~Vernizzi,
  Phys.\ Rev.\ Lett.\  {\bf 95}, 091303 (2005)
  [arXiv:astro-ph/0503416];
  {\em ibid.}, Phys.\ Rev.\ D {\bf 72}, 103501 (2005)
  [arXiv:astro-ph/0509078].



\bibitem{Langlois:2006iq}
  D.~Langlois and F.~Vernizzi,
  JCAP {\bf 0602}, 014 (2006)
  [arXiv:astro-ph/0601271].


\bibitem{Langlois:2006vv}
  D.~Langlois and F.~Vernizzi,
  arXiv:astro-ph/0610064.

\bibitem{Bardeen}
  J.~M.~Bardeen, P.~J.~Steinhardt and M.~S.~Turner,
  Phys.\ Rev.\ D {\bf 28} (1983) 679.



\bibitem{Hwang:1991aj}
  J.~Hwang,
  Astrophys.\ J.\  {\bf 375}, 443 (1991).

\bibitem{Wands:2000dp}
  D.~Wands, K.~A.~Malik, D.~H.~Lyth and A.~R.~Liddle,
  Phys.\ Rev.\ D {\bf 62} 043527 (2000)
  [arXiv:astro-ph/0003278].


\bibitem{Malik:2003mv}
  K.~A.~Malik and D.~Wands,
  Class.\ Quant.\ Grav.\  {\bf 21}, L65 (2004)
  [arXiv:astro-ph/0307055].

\bibitem{ellis}
  G.~F.~R.~Ellis, Relativistic Cosmology, in {\em General Relativity
    and Cosmology}, proceedings of the XLVII Enrico Fermi Summer School in
  Varenna, Italy, 1969, edited by R.~K.~Sachs (Academic, New York, 1971).

\bibitem{Stewart:1977}
  J.~M.~Stewart and M.~Walker,
  Proc.\ R.\ Soc.\ London  {\bf A341}, 49 (1974).


\end{thebibliography}
\end{document}